ARTICLE

# XR and Hybrid Data Visualization Spaces for Enhanced Data Analytics

Santiago Lombeyda[1] | S. G. Djorgovski[1] | Ciro Donalek[1,2]

[1]*Center for Data-Driven Discovery, Division of Physics, Mathematics, and Astronomy, California Institute of Technology, Pasadena, CA 91125, USA*
[2]*Virtualitics, Inc., 225 S. Lake Avenue, Suite 120, Pasadena, CA 91101, USA*

**Correspondence:** Santiago Lombeyda, MS 158-79, Caltech, Pasadena, CA 91125, USA (santiago@caltech.edu)

**ABSTRACT**

The growing complexity and information content of data, together with the need to understand both the complex structures, relationships, and phenomena present in these data spaces, compounded with the emerging need to understand the results produced by AI tools used to analyze the data, requires development of novel, effective data visualization tools.  Much of the growing complexity is reflected in the increasing dimensionality of data spaces, where extended reality (XR) naturally emerges as a candidate to help extend our capability for higher dimensional understanding. However, humans often understand lower dimensionality representations more effectively.  Still, XR offers an opportunity for a seamless integration of simulated traditional data displays within the 3-dimensional virtual data spaces, leading to more intuitive and more effective data analytics. In this paper we present an overview of the benefits of seamlessly integrated 2-dimensional and 3-dimensional interactive visual representations embedded in XR spaces, and present three case studies that leverage these approaches for more efficient data analytics.

## 1 | INTRODUCTION

Effective data visualization is a critical, if often underestimated, component at every stage of the data-into-knowledge process, from the initial data inspection and cleaning, through the choices of data analysis tools and algorithms, understanding and interpretation of their output, to the presentation and publication of results.

The core purpose of data analytics is the discovery and interpretation of meaningful patterns that may be present in the data. Visualization connects the quantitative content of data with human intuition and understanding, and it has been said that we cannot really understand or intuitively comprehend anything, including mathematical and other abstract concepts, that we cannot visualize in some way.  Human pattern recognition and other cognitive processes are often based on the visual information channel.  Effective data visualization is thus one of the key methodological challenges in the era of exponential data growth.  As we increasingly use Artificial Intelligence (AI) tools and depend on them for data analysis and data-driven decision making, visualization is also a key component of explainable AI.  Put simply, visualization is at the intersection of big data, machine intelligence, and human understanding.

Data analytics challenges include rapidly growing data rates and volumes, but perhaps the most difficult come from the increase in data complexity and information content. Data complexity may be expressed through the dimensionality of data feature spaces (number of independent variables captured in the data and their interaction) and data heterogeneity (variety of data types).  Data complexity imposes challenges for both the scalability of Machine Learning (ML) algorithms and for data visualization.  Visualizing and analyzing 1-Dimensional (1-D; e.g., time series, probability density distributions) or 2-Dimensional data (2-D; e.g., images, maps, simple scatter plots) is readily done with the common visualization tools.  But if the data populate some manifold with a higher dimensionality $D > 3$, projecting them to a lower dimensionality visual space inevitably involves a loss of information.

How do we visualize data spaces with many more than 3 dimensions?  This is where judiciously used eXtended Reality (XR) can help, and we describe some examples in Figure 1.  Numerous studies in a variety of fields have used VR, and found it superior to the traditional (flat) representations, at least for some tasks; see, e.g., [1] as an example.

XR is not a panacea; it can help in some situations, e.g., with the data spaces of inherently high dimensionality, or even just 3-D data where the 3-D context is important.  Yet, most experts are trained in the use of 1-D or 2-D data representations.  Here we offer examples of successful data visualization in XR, but also argue that in some situations hybrid approaches using XR and 1-D and/or 2-D modalities may yield better results.  Another advantage of XR is that it facilitates an intuitive collaborative data analysis, where the researchers, their data, and computing tools, all of which can be geographically distributed, share a common virtual workspace.

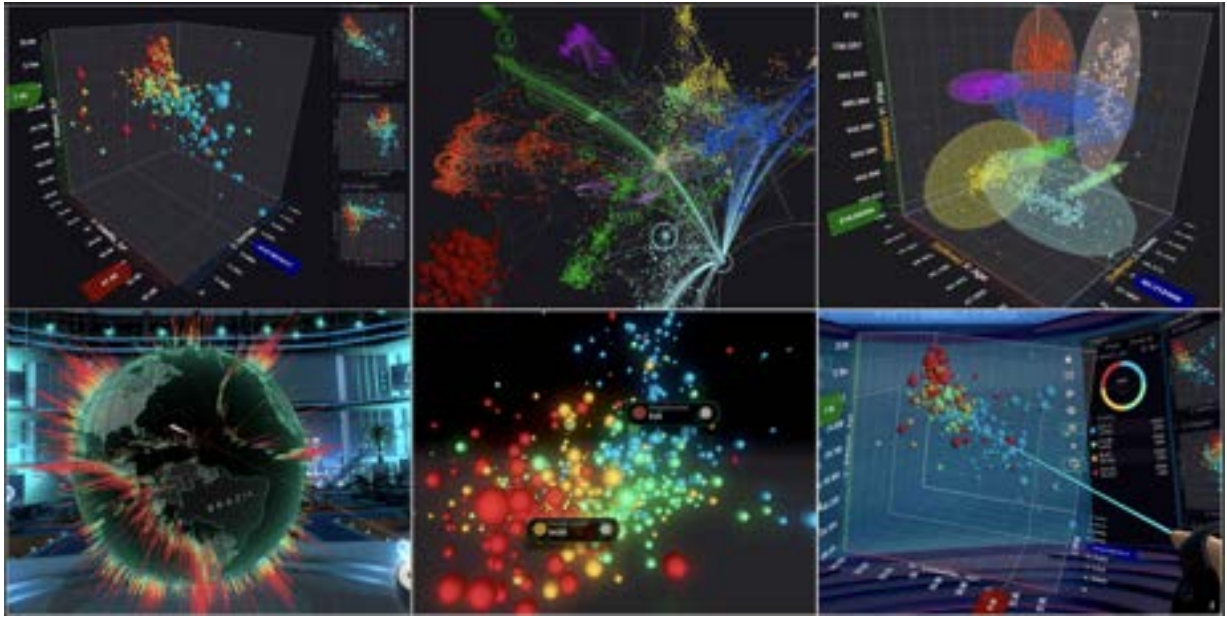

**Figure 1** | Examples of a variety of interactive data analysis in a Virtual Reality environment. Images courtesy of Virtualitics, Inc.

## 2 | THE ORIGINS OF XR

The introduction of Virtual Reality (VR) headsets in 1965 [2] that was followed by the first commercial headsets in the 1980s [3], provided a glimpse of what it would be like to be immersed in a fully virtual environment. From the infancy of VR, it was apparent that the experience of being in a virtual or augmented world could enable users to accomplish extremely complex tasks in intuitive new ways [4], including data immersion and analysis. But the larger promises of seamless data inspection spaces were often relegated to niche research into the spaces themselves, proof of concept prototypes often employing overcomplex layouts that were only helpful for small spatial tasks, and science fiction movies.

As VR has led the way into Augmented Reality (AR), we now understand that when we pair visuals and spaces alongside additional sensory feedback [5], including haptics, we increase the sense of presence, which clears the way for our intellectual system to believe we are physically present alongside these virtual constructs. This creates a stronger sense of immersion that fosters a more effective space for our cognitive system to scaffold information, retain physical constructs [6], and have a richer understanding of physical structures [7], which can ultimately create more efficient analysis tools.

The introduction of the *Oculus Rift* in 2012 [8] kickstarted the new wave of hardware innovations, with VR products from different companies providing increases in screen definition and field of view. Remote work needs during the Covid-19 pandemic (2020-2022) accelerated the need for virtual office environments with private (virtual) screens. Industry attempted to take advantage of this need (for high-quality 2D spaces in XR) to consolidate a viable market that resulted in the introduction of a wide variety of VR and XR hardware aimed at productivity and business needs. In the end, display quality, complexity in the interactivity, and the form factor of the tools themselves were ultimately not conducive to allowing this market to flourish. The persistence creativity and ambition pushing for innovation in the field, and the drive to match hardware and potential markets, alongside synergy with the boom of Artificial Intelligence (AI), have meant that the XR field continues to be vibrant, hardware continues to be created, and industry is actively looking for better experiences, better interfaces, and ultimately good use of their products.

As of mid-2025, commercially available XR hardware includes the *Oculus Quest 3* [9], the *Apple Vision Pro* [10], as well as a wide variety of *Snapdragon*-powered AR glasses like the *XREAL Pro One* [11]. Commodity consumer and enterprise hardware is now widely affordable for an average user, with options ranging from connecting to a desktop or laptop, running a mobile device or portable computing and power brick (also referred to as a puck), or working independently or directly from the cloud.

## 3 | VIRTUAL 3-D SPACES

At the most elemental, we live our daily lives in a perceived physical 3-D universe, filled with 3-D phenomena. Understanding these 3-D physical relationships requires spatial understanding that needs to thread information between three-dimensional representations and descriptive one-dimensional (1-D) or two-dimensional (2-D) representations. Two-dimensional visualization of abstract information can be superior to 3-D counterparts [12,13]. However, as a large number of datasets map to 3-D structures, those could be easier to understand in the full three dimensions over 2-D flattened versions. For instance, a study in measuring completion time of tasks and error margins for a radiology case study, shows that while experts prefer 2D representations, error is reduced with 3D and 2D/3D hybrid solutions, while 3D is in fact more efficient for non-experts [36]. Of course, added complexity can again render straightforward 3-D representations of limited utility, and higher-dimensional data can easily overwhelm straightforward 3-D mappings and not reveal structures embedded in the higher dimensions.

Whereas virtual 3-D spaces are at least the first obvious, strong, and viable methodology to begin to showcase 3-D (spatial) relationships, even if ultimately our brain will need to decode the information into lower-dimensional abstractions [14]. Visual exploration solutions can then supplement the deeper understanding and analysis mechanisms through complementary and connected visual paradigms, lots of which are already part of our scientific exploration toolkits.

The effectiveness of 3-D spaces over 2-D spaces, or 3-D representations over 2-D representations, depends on the task. If the actionable result is intrinsically connected to the representation, 3-D representations will more likely result in increased successes (such as in [15] or [16]). If the actionable outcome is more dependent on abstraction or summarization of perceived arrangement, then 2-D representations are likely to be more effective [17]. Figuring out the space projection given the task can be quite a hard problem, as even the more basic rules of perception when applied to visualization are rarely followed [18].

## 4 | BEYOND 3-D SPACES

Interactive virtual 3-D space representations span from the fully realistic or augmented realistic spaces to completely abstract creations. All of them can engage our senses and ultimately feel real, especially when they create cognitively harmonious connections with our physical, natural real spaces [5].

In general, there is not a single universal representation that would allow us to understand complex relationships that may be present in data. We need to utilize all the tools at our disposal to be able to contextualize information, create mechanisms to represent complexity and highlight detail, and ultimately be able to tabulate the encoded relationships, both those that are exposed on visual exploration, as well as those that we cannot see easily. Data visualization should utilize as many representations as needed to be able to elucidate the patterns that need to be recognized.

It is impossible to generate a universal tool or environment that will generally address every researcher's data analysis needs. However, we serve as example particular case scenarios where (a) complex real science can make strong use of 3-D spaces, (b) address how 2-D representations can be embedded and connected into a cohesive immersive environment, and (c) offer the conceptual template on how the current available technologies can be utilized to recreate and adapt these solutions for similar scientific data analysis endeavors.

The use of XR, where a user can view the real world through specialized XR headsets or XR glasses offers users the ability to be present in their usual work environment, while integrating virtual 2-D and 3-D interactive representations, integrated in the same virtual data space. Furthermore, having direct access to work-proven classic input peripherals (keyboard and mice) alongside their usual computing environment, whether through an actual desktop or notebook with a physical display or through a virtual 2-D screen, enables users to seamlessly continue their usual workflows, while their classic tools coexist with in-situ 3-D interactive representations. Connecting the 3-D models and the 2-D interfaces and representation would thus only require a common communication interface to allow cross-tool interaction. Through the use of the standard communication sockets, this is now readily accomplishable. Tools and interaction can be designed to work across different tools, users, and data, as was presented in [19]. Meanwhile, the way 2-D and 3-D representations coexist and interact in meaningful efficient ways (Figure 2) has also been studied, such is the work presented in [20].

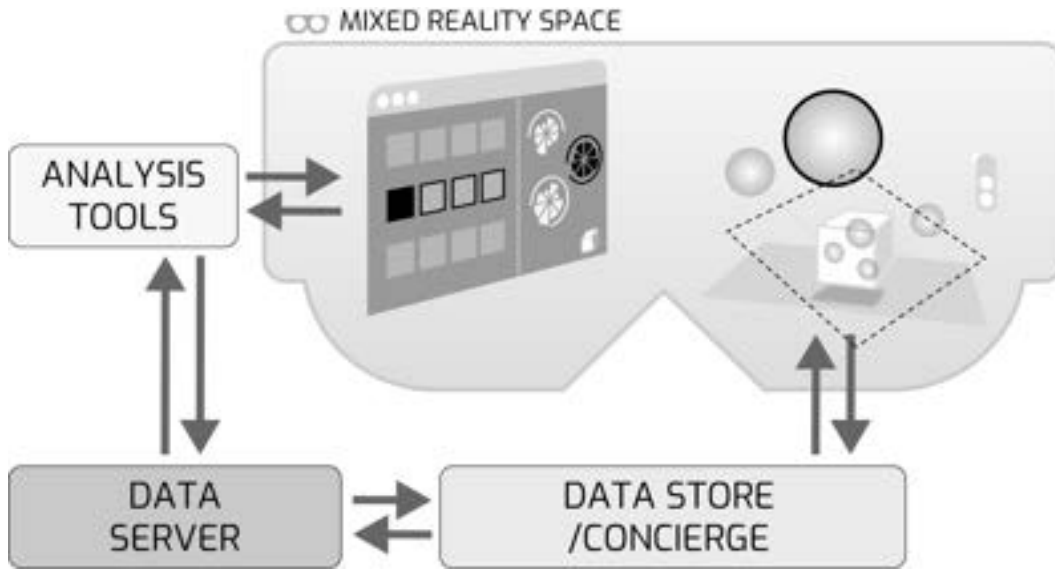

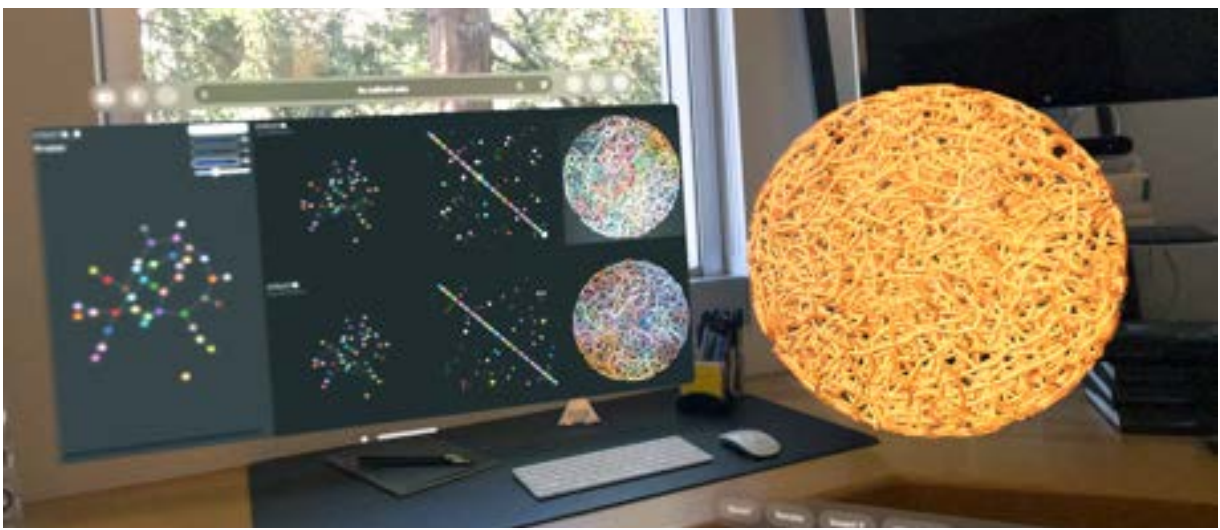

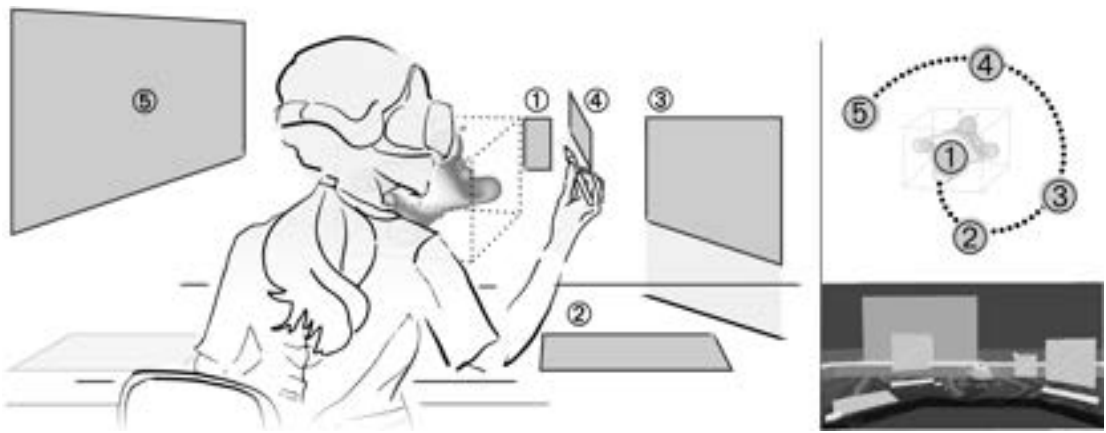

**Figure 2 | (Top)** Illustration of how 2-D screen based tools and 3-D in space representations can interact, with data server and analysis tools working together. **(Middle)** Screenshot from extended reality work session using Apple's Vision Pro, connecting a screen based 3D-DNA interactive visualization portal, alongside actual structure embedded into the workspace. **(Bottom)** A conceptual illustration of an XR desktop-based workspace that can be used to contextualize data at different scales and regions of focus, alongside informational and fully interactive 2-D screens embedded in 3-D space [19].

## 4.1 | XR+AI

In recent years, Generative AI has emerged as a valuable complement to both traditional and immersive approaches, from automating visualization creation to enhancing data exploration and storytelling. For example, recent agentic systems are used for interactive text to visualization, providing the capability to create and manipulate 3D models, in real time, by voice [37].

Generative models can produce contextually relevant visual, textual, and multimodal representations that bridge complex, high-dimensional data spaces and human cognition. When coupled with XR environments, these capabilities can dynamically adapt visual encodings in real time, generate explanatory annotations, or create immersive narratives that guide users through their data.

Several emergent platforms and research prototypes illustrate how generative models are enhancing XR-driven data analytics. From AI-powered storytelling engines [21] to content generation [22] and open-source XR engines [23], the field is converging toward dynamic, responsive, and intuitive immersive experiences for data exploration. For instance, generative AI has been integrated into hybrid mixed-reality workflows to streamline collaborative forensic autopsy [38]. Moreover, open platforms for collaborative 3D simulation such as NVIDIA Omniverse support XR-accessible digital twins and integrate generative AI microservices, enabling rapid authoring and interactive immersion of digital-twin environments [39].

## 4.2 | Case Study: 3-D DNA

Molecular biologists understand that both the structure and relative physical location of Chromosomes inside a cell serve to facilitate communication (expression) and allow a cell to carry out a variety of different functions. Understanding the genome structure is an area of active research. This includes understanding the organization into chromosome territories, how DNA interaction among the same chromosome can happen through chromatin loops, leading to understanding local-interacting neighborhoods called "topologically associated domains" (TADs) [24,25].

Physical structure of DNA is revealed through chemical processes, which break apart these long molecular chains, and then rebuild them, giving them the ability to extrapolate the original structure and physical packing of chromosomes inside a cell for a given particular setup. While 3-D representations were needed to gain first insight into the structures and the physical relationships between the different segments, just visualizing the 3-D structures was not very useful. The first goal for a visualization tool arose from the need to understand how the different sections of chromosomes (TADs) were in "communication" with each other.

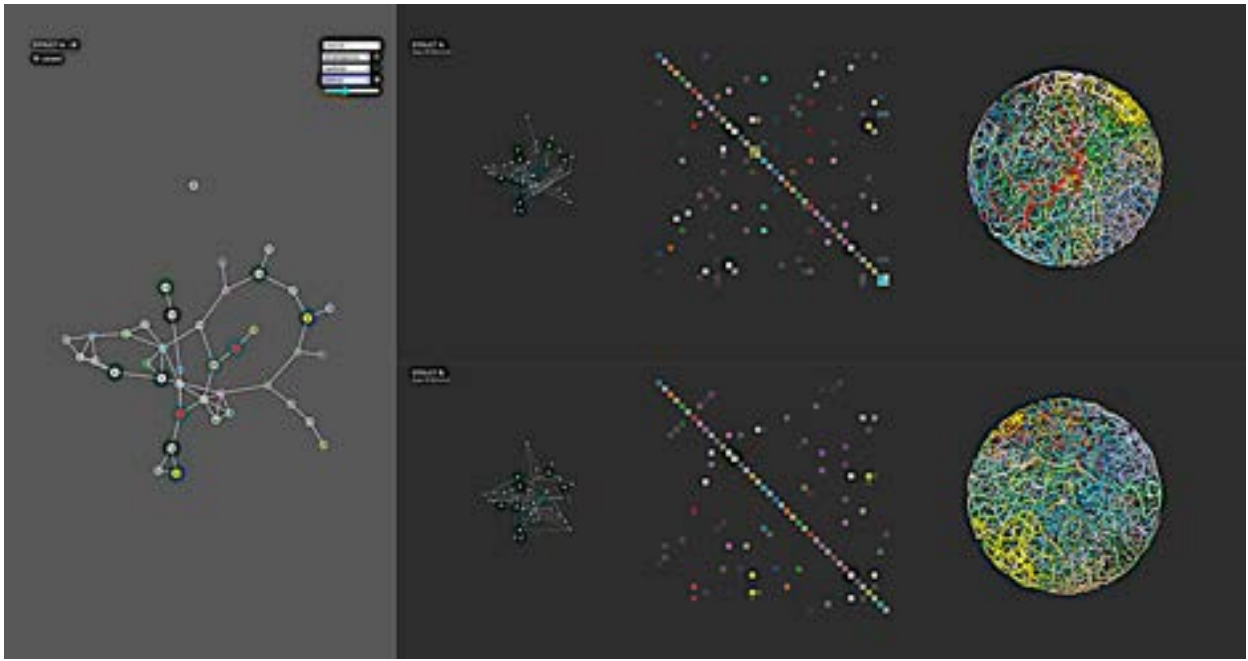

**Figure 3** | Screenshot of web-based portal used to visualize, explore, and compare 3-D DNA structure inside a cell, with connected views showing (from right to left) the inferred 3-D arrangement of chromosomes, an interaction matrix, and individual and global connectivity graphs. Work from JPL/Caltech/ArtCenter's Data to Discovery program in collaboration with Guttman Lab, Caltech.

Therefore, representations well suited for this task included relationship matrices and relationship graphs, with different segments or TADs represented as rows and columns in the matrix, or nodes in a network graph (Figure 3). In order to help researchers make sense of both the structures and the sequence of connections, we placed these representations next to each other and enabled interactivity through "brushing". Users were able to easily select segments in any of the displays, and the tool would highlight all related sections and neighborhoods across all plots. The tool also displayed the DNA structure in 3-D by itself. However, the 2-D flattened representation on a 2-D screen was nearly impossible to understand. Interactivity allowed the structure to be rotated. When in motion, or simply spinning, users could start to understand some of the 3-D spatial relationships. Yet, it was extremely difficult to create a memory impression of where different structures were relative to each other, especially as the objects were in a constant motion; when the movement or spinning was halted, there was an almost immediate loss of the 3-D spatial relationships. It was when these representations were placed in the VR that we could keep this object static while preserving the cognitive understanding of 3-D relationships.

The interactivity of the 2-D representations was key in starting to understand the richer relationships, but the context of the 3-D structure was key not only for the original physical setup of the phenomena, but to justify the dependencies and interactions among the different segments. As a result, a 3-D (seemingly) static representation in XR serves as a much stronger contextual and navigational representation than a 3-D representation in motion flattened to a 2-D screen.

This project was originally deployed in 2018. At that point, 2D interfaces inside VR were possible, but were far too low resolution to create a fully usable 2-D plus 3-D immersive work space.  However, the prototype in VR validated our proposal that intuitive understanding of 3-D DNA structures in VR would allow mental models of relationships and spatiality to co-exist with the analysis 2-D interfaces, without the need for the 3D object to spin or have continuous artificial movement, as was needed in 2-D.

The 3D-DNA tool developed in 2018 runs seamlessly via supported browser windows, including in XR (Figure 2). Meanwhile, the 3-D DNA model import into the virtual space was also a very straightforward task of exporting each data segment (chromosome) as 3-D lines in space that were then extruded into tubes at runtime, or alternatively, exported directly as meshes.

### 4.3 | Case Study: Lung Tumor + ML

ML and AI have a substantial potential to accelerate cancer tumor detection, including an unprecedented ability to scan patient data and find vulnerable areas and catch nascent cancer cells before it is even visible to humans studying the different data points. However, it all starts from the creation and curation of training datasets, where experts, clinicians, oncologists, and radiologists have gone through, tagged, measured, qualified, and classified tumors from patient scans.  The larger the training data set is, the higher the precision of the results.

Currently, clinicians are trained on 2-D scans, where the location and orientation of the scan allow them to discern anomalies, in the context of organs and key organic structures.  In order to create a better visual context of the data, we created OVS+Tumor, a tool in VR where the user can explore a real patient scan, home into an area of interest, and create 3-D structures (isosurfaces) from the scans. We narrowed our interest to the lung structures, where bronchioles are quite difficult to understand from 2-D slices. However, creating unfamiliar 3-D representations for clinicians felt unfounded and hard to understand. [26] embedded the 2-D slices into the volumetric representation, allowing easy toggle between 2-D and 3-D and coexisting representations.

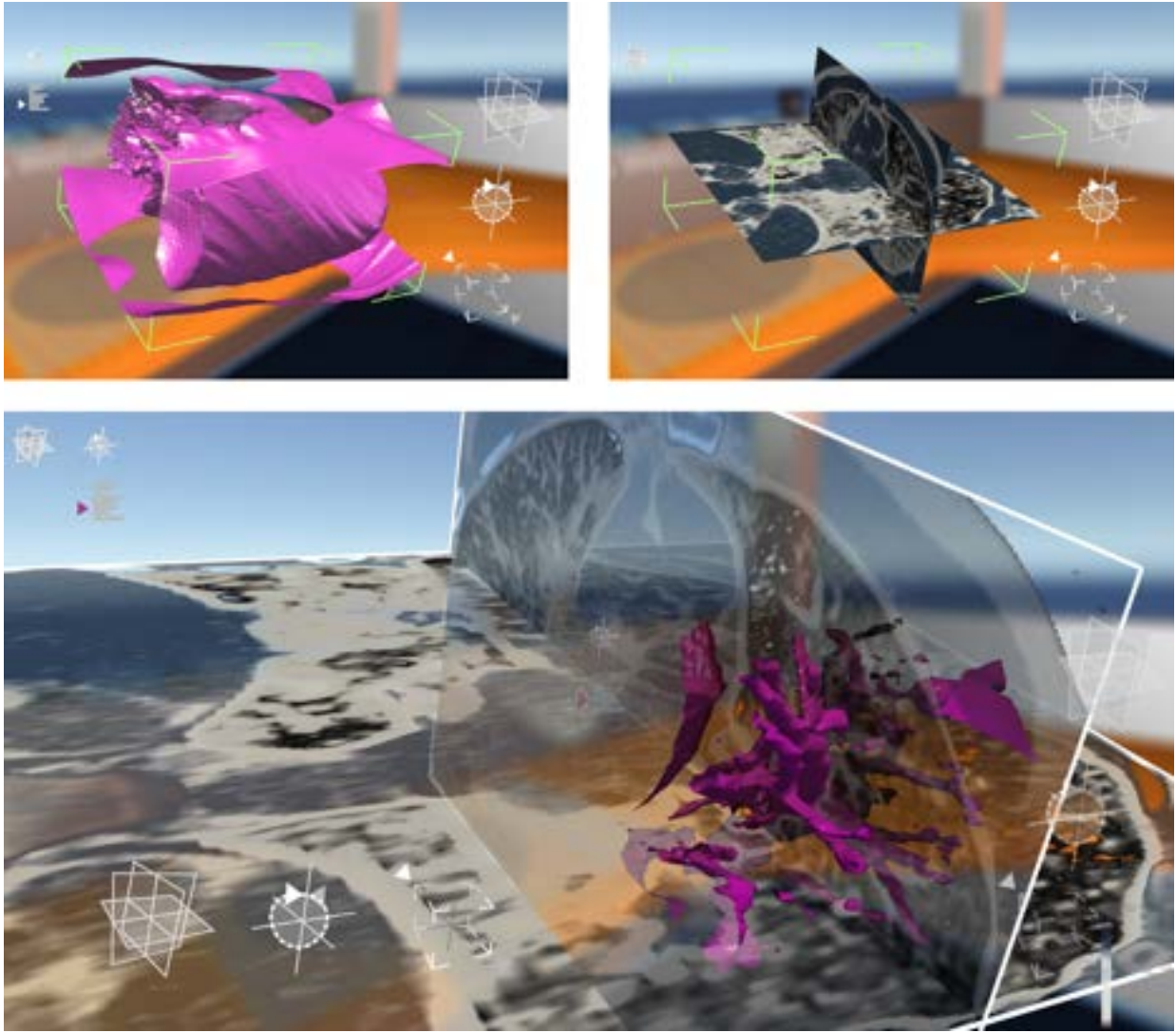

**Figure 4** | Example of 2-D CT scan used in context with 3-D surface extraction, as part of OVS+TUMOR interactive tumor visual classification tool created in VR [26].

The creation of such a tool (Figure 4), while creating a much simpler entry point for non-experts, for experts it allowed better contextualization of the information, easier verification of information across slices (sequential images), and ultimately it allowed them to incorporate a new way of understanding their data, in a space that felt comfortable and efficient. OVS+Tumor allowed for direct manipulation of the volume (3-D structure) and easy markup of regions or structures of interest through simple selection of objects, or through the use of an in-space 3-D pen tool. OVS+Tumor was implemented fully using *Unity* [27], while utilizing extensions for embedded browsers to display additional 2-D information, as well as in-space embedding for 2-D plates collocated with 3-D structures.

While OVS+Tumor proved to be efficient and intuitive, more nuanced marking and measurement of tumors could be more easily done by moving back and forth between direct volume interaction and analysis of in-space 2-D representations. Marking the main axis of a tumor in a 2-D slice could be seen as the closest view of truth in the data that clinicians have, while the 3-D representation (and projected axis representation and measurement) can serve to verify the physicality of the found tumor and relationship to the mass around it.

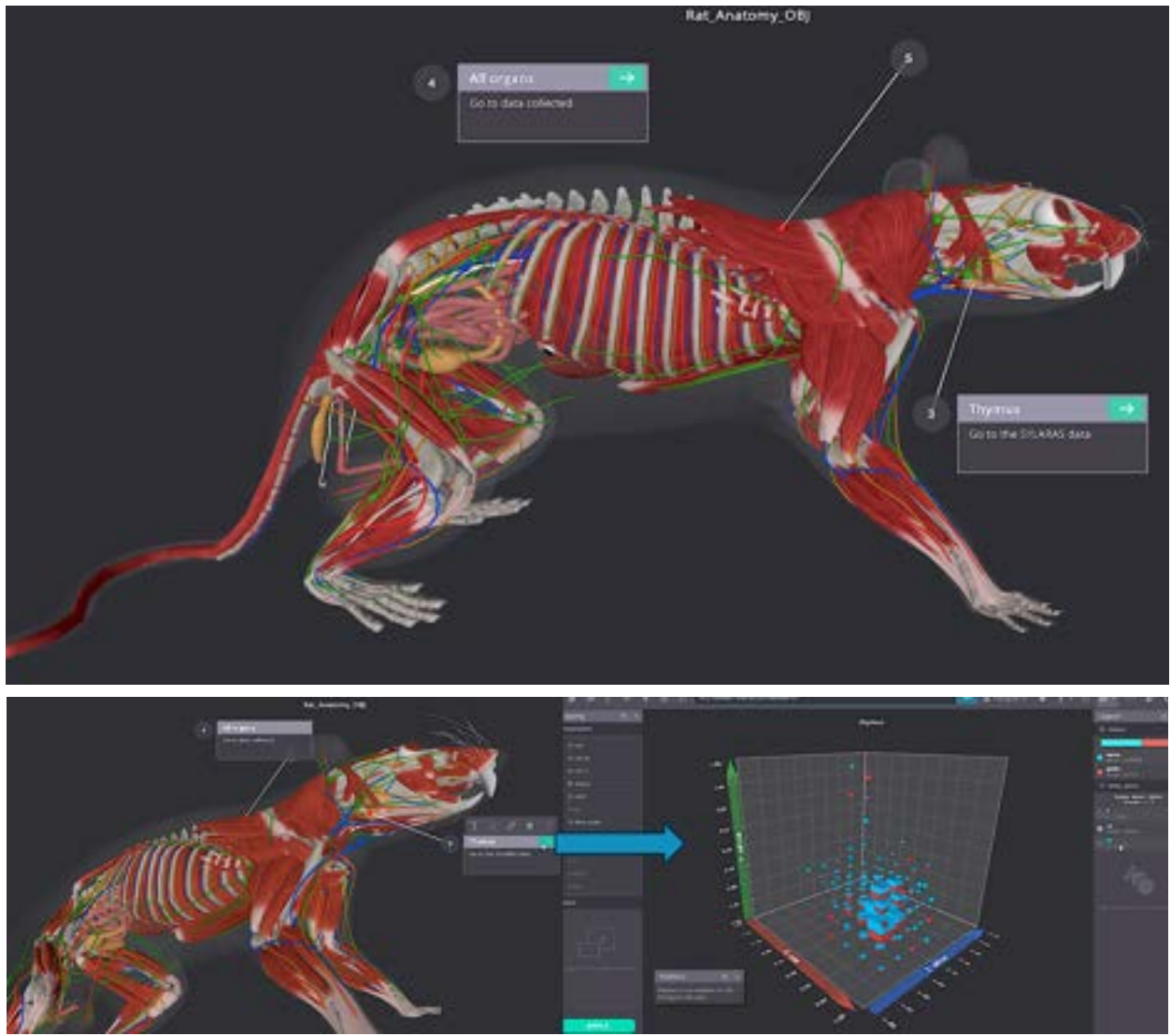

**Figure 5** | Inferring Network-level Immune Interactions using preclinical cancer models and AI-driven data analytics and visualization. Images courtesy of Virtualitics, Inc.

### 4.4 | Case Study: 3D preclinical cancer model + AI-driven data analytics

Extended Reality (XR) is reshaping how we interact with complex biomedical data, moving from static charts and panels to immersive, intuitive environments. In this proof of concept, a 3D preclinical mouse model was created, with each organ annotated and connected to underlying datasets [28,29]. By simply selecting an annotation, researchers could jump directly into AI-driven analyses and begin exploring results in real time. This transforms the traditional dashboard into an immersive space where scientists can "walk through" their data, making exploration more natural and grounded in the biological context (Figure 5).

This shows that the potential of XR goes far beyond visualization. By combining immersive models with advanced analytics and narrative design, we can turn data into stories that connect computational insights with experimental expertise. In this case, the 3D mouse model served as both a biological reference and a dynamic interface to infer network-level immune interactions in cancer research. This approach illustrates how XR can accelerate discovery, enable new forms of collaboration, and make data actionable offering a pathway to more informed decision-making and a deeper integration of AI into scientific practice.

## 5 XR AND HYBRID SPACES

These are some of many examples that we have seen where 3-D in VR or XR is used to create navigational context that is crucial for us to be able to understand the breakdown of all the data. However, what is still missing is the ability to create a flow of data from representation to selection to enable further detailed analysis. It is through manipulation in conjunction with computing and analysis that we can foment discovery. Such is only made possible if we have tools at our disposal that can be then interconnected and exposed as needed. For our scientific studies, that usually has meant elements including a console running Python, a Jupyter notebook, tools developed in Julia or Rust, as well as tools and modules compiled in different languages.

It is with the promise and now existence of XR spaces that treat 2-D visual displays (i.e., virtual screens and windows) rendered in high resolution as important elements of the virtual environment, that give the ability to see details and read data from within our XR spaces. Adding a visible keyboard and being able to bring all of the desktop tools onto the virtual desktop space allows XR spaces to fulfill the promise of fully connected representations for computation and analysis.

### 5.1 FROM WEBXR TO ANDROIDXR

The largest obstacle in creating seamless tools will be the ability to enable current workflows, and then easily extend those to be part of a larger software ecosystem that connects with exterior representations and compute resources, and allows for easy flow of communication among them. Game development environments, such as *Unity* [30] have embraced XR and fully support different levels of hardware, with different levels of API abstraction. The ability for these game development environments is facilitated by consortia like W3C [31]. W3C's support for WebGL [32] and WebGPU [33] is the backbone of interactive graphics on the web and portable graphics applications across different devices.

Similarly, leveraging WebXr [34], AndroidXR [35] proposes to create an open platform for XR, supporting multiple hardware, peripherals, and a wide set of modes of operation. As the collaboration grows between chipmaker SnapDragon and Google (makers of AndroidXR), the expectation is that AndroidXR may serve as a possible strong development space to more easily create these integrated applications.

## 6  CONCLUDING COMMENTS

The growing need for more effective data visualization is driven largely by the growing complexity of modern data sets that often manifests as an increased dimensionality. Visualizing such data spaces in a way that facilitates discovery and intuitive understanding of complex relationships present in the data requires novel approaches, including XR.  However, there is an equally important and growing need to understand the results of ML/AI analysis of such data spaces, i.e., the explainable AI (XAI).  Both of them call for the visualization solutions that combine new display technologies and a better, user-centric design of data exploration and analytics systems.

One important channel of communication that is currently often ignored in data exploration is the back-and-forth analysis between 2-D data/representations with their 3-D phenomena context from which the data has originated.  XR is not a panacea: 3-D representations are often burdened with having to answer too many questions, and, given that historically creating 3-D models and then rendering them in any kind of interactive way has been computationally difficult, we tend to shy away from these. The answer to this problem is through thoughtful creation of blended environments that allow our data to exist across 2-D panels and 3-D in-space representations, with these 3-D objects serving as a central navigation map and context.

A full understanding of the spatial relationships, while allowing for an efficient 2-D window-based interaction tool is only possible through XR immersive spaces. Creating seamless tools can only happen through APIs that easily support 2-D panels alongside 3-D models and allow for easy communication back and forth.

The current hardware is good enough to do this. The software exists and is slowly getting easier to create these connected tools, across most hardware. Interaction mechanisms are not fully developed or feel seamless.  Keyboards and mice that interact with 2-D panels are now fully supported with new hardware and software APIs.  Gesture and voice recognition is still imperfect, but circumventable for now, and steadily improving.

We envision that a new generation of data analytics systems would require designs that address and facilitate the three-way interactions between users, their data, and AI tools, with user interactions that include visual channels, haptics, and natural language processing.  New discoveries may be hiding in the data because we cannot see them yet.

**ACKNOWLEDGMENTS**

SL and SGD acknowledge the generous support for this work from Ajax Foundation. We are grateful for the partnership and collaboration with Caltech's Guttman Lab, JPL and the NCI's Early Detection Research Network, Caltech's OVRAS Lab, and Caltech/JPL/ArtCenter Data to Discovery program.